\documentclass[%
 reprint,
 amsmath,amssymb,
 superscriptaddress,
 aps
]{revtex4-2}

\usepackage{graphicx} 
\usepackage{dcolumn} 
\usepackage{bm} 
\usepackage{enumerate}
\usepackage{mathtools}
\usepackage{physics}

\makeatletter
\let\MYcaption\@makecaption
\makeatother
\usepackage{subcaption}
\makeatletter
\let\@makecaption\MYcaption
\makeatother

\usepackage{hyperref}

\begin{document}


\title{Visualizing convolutional neural network for classifying gravitational waves from core-collapse supernovae}

\author{Seiya Sasaoka}
\affiliation{Department of Physics, Tokyo Institute of Technology, 2-12-1 Ookayama, Meguro-ku, Tokyo 152-8551, Japan}%

\author{Naoki Koyama}
\affiliation{Graduate School of Science and Technology, Niigata University, 8050 Ikarashi-2-no-cho, Nishi-ku, Niigata City, Niigata 950-2181, Japan}%

\author{Diego Dominguez}
\affiliation{Department of Physics, Tokyo Institute of Technology, 2-12-1 Ookayama, Meguro-ku, Tokyo 152-8551, Japan}%

\author{Yusuke Sakai}
\affiliation{Department of Design and Data Science and Research Center for Space Science, Advanced Research Laboratories, \\ Tokyo City University, 3-3-1 Ushikubo-Nishi, Tsuzuki-ku, Yokohama, Kanagawa 224-8551, Japan}%

\author{Kentaro Somiya}
\affiliation{Department of Physics, Tokyo Institute of Technology, 2-12-1 Ookayama, Meguro-ku, Tokyo 152-8551, Japan}%

\author{Yuto Omae}
\affiliation{Artificial Intelligence Research Center, College of Industrial Technology, Nihon University, 1-2-1 Izumi-cho, Narashino, Chiba 275-8575, Japan}%

\author{Hirotaka Takahashi}
\affiliation{Department of Design and Data Science and Research Center for Space Science, Advanced Research Laboratories, \\ Tokyo City University, 3-3-1 Ushikubo-Nishi, Tsuzuki-ku, Yokohama, Kanagawa 224-8551, Japan}%
\affiliation{Institute for Cosmic Ray Research (ICRR), The University of Tokyo, 5-1-5 Kashiwa-no-Ha, Kashiwa City, Chiba 277-8582, Japan}
\affiliation{Earthquake Research Institute, The University of Tokyo, 1-1-1 Yayoi, Bunkyo-ku, Tokyo 113-0032, Japan}

\date{\today}

\begin{abstract}
In this study, we employ a CNN to classify GWs originating from CCSNe. Training is conducted using spectrograms derived from three-dimensional numerical simulations of waveforms, which are injected onto real noise data from the O3 observation of both Advanced LIGO and Advanced Virgo. To gain insights into the decision-making process of the model, we apply CAM techniques to visualize the regions in the input image that are significant for the model’s prediction. The CAM maps reveal that the model’s predictions predominantly rely on specific features within the input spectrograms, namely the $g$-mode and low-frequency modes. The visualization of convolutional neural network models provides interpretability to enhance their reliability and offers guidance for improving detection efficiency.
\end{abstract}


\maketitle

\section{Introduction\label{sec:intro}}
The first detection of GWs from a binary black hole merger by Advanced LIGO~\cite{ref:Aasi2015} in 2015 marked the beginning of GW astronomy~\cite{ref:Abbott2016}. Throughout three observing runs (O1, O2, and O3), Advanced LIGO and Advanced Virgo~\cite{ref:Acernese2015} reported 90 GW events~\cite{ref:Abbott2019, ref:Abbott2020, ref:Abbott2021_a, ref:Abbott2021_b}. As of May 2023, the international GW network, now including KAGRA~\cite{ref:Akutsu2019}, has begun its fourth observing run (O4) with improved sensitivity. 

All of the GW events detected so far are exclusively from compact binary coalescences. However, short-duration GW bursts arising from CCSNe are expected to be detected by the current and next-generation GW detectors, such as the Einstein Telescope~\cite{ref:Punturo2010} and the Cosmic Explorer~\cite{ref:Abbott2017_ce}. CCSNe, resulting from massive star explosions leading to neutron stars or stellar-mass black holes, stand as one of the most energetic astrophysical events in the Universe, emitting electromagnetic waves, neutrinos, and GWs. While electromagnetic waves from CCSNe are frequently observed, neutrinos have only been detected from SN1987A~\cite{ref:Bionta1987, ref:Hirata1987}. GWs are expected to carry information about the inner core's dynamics, providing vital insights into the explosion mechanism which remains elusive. The primary conundrum lies in discerning how a stalled shock wave is revived to cause a star to explode. Currently, there are two prevailing theories~\cite{ref:Janka2012}: the neutrino-driven mechanism~\cite{ref:Bethe1985}, in which shock waves are revived by neutrinos stored behind the shock wave heating the surrounding matter, and the magnetorotational mechanism~\cite{ref:LeBlanc1970}, in which the rapid rotation of the progenitor causes explosions driven by strong magnetic fields. The typical GW detection range for neutrino-driven signals is expected to be around 10 kpc, while the detection range for magnetorotational signals is expected to be above 100 kpc~\cite{ref:Marek2021}.

Due to the stochastic nature of GW signals from CCSNe, the conventional matched-filtering technique, which relies on specific waveform templates, is unsuitable. Alternative detection methods based on the time-frequency representation have been devised in response. In particular, the coherent WaveBurst (cWB) pipeline~\cite{ref:Klimenko2004, ref:Klimenko2016} detects and reconstructs burst GW signals by searching for excess power in a time-frequency map, with minimal reliance on a specific source model.

Predicting GW signals from CCSNe remains a formidable challenge. However, recent advancements in theoretical research and multidimensional numerical simulations have revealed certain signal properties. For neutrino-driven CCSNe, the dominant emissions arise from the $g$-mode oscillation of the proto-neutron star (PNS) surface. These frequencies progressively increase over time, ranging from a few hundred Hz to a few kHz. Additionally, at low frequencies ($\lesssim$ 200 Hz), GW emissions associated with hydrodynamics instabilities including neutrino-driven convection and standing accretion shock instability (SASI)~\cite{ref:Blondin2003} are observed in some simulations. Insights obtained from these simulations are pivotal in enhancing methods for CCSNe detection and analysis.

In recent years, machine learning techniques, especially deep learning, have gained traction in a variety of scientific fields due to their capacity for recognizing intricate patterns and extracting meaningful features from large data sets. This ability has been especially noted in areas such as computer vision and natural language processing. Its application in GW research has followed, with numerous implementations and explorations, as highlighted in a comprehensive review in Ref.~\cite{ref:Cuoco2020} and the foundational efforts by \citeauthor{ref:George2018}~\cite{ref:George2018, ref:George2018_2}. In the field of CCSNe analysis, \citeauthor{ref:Astone2018}~\cite{ref:Astone2018} leveraged CNNs to detect CCSNe within Gaussian noise, using $g$-mode phenomenological waveforms for training, outperforming the cWB pipeline. Consecutive studies by \citeauthor{ref:Iess2020}~\cite{ref:Iess2020, ref:Iess2023} involved the training of both one- and two-dimensional CNNs and long short-term memory networks~\cite{ref:Hochreiter1997} to identify seven distinct CCSN waveforms embedded in real noise and glitches, with their models achieving 98\% classification accuracy at 1 kpc with a three-detector network. Additionally, \citeauthor{ref:Chan2020}~\cite{ref:Chan2020} employed one-dimensional CNNs to investigate both magnetorotational and neutrino-driven signals in Gaussian noise, recording a true alarm probability of 80\% for magnetorotational signals from sources at 60 kpc and 55\% for neutrino-driven signals from sources at 10 kpc with a fixed false alarm probability of 10\%. In another study, \citeauthor{ref:Edwards2021}~\cite{ref:Edwards2021} used two-dimensional CNNs to classify 18 different equations of state (EOS) from pure magnetorotational CCSN signals, attaining an accuracy of 72\%. \citeauthor{ref:Lopez2021}~\cite{ref:Lopez2021} refined phenomenological waveforms originally used by \citeauthor{ref:Astone2018}~\cite{ref:Astone2018}, achieving a 60\% true alarm probability for signals located at 15 kpc with a 5\% false alarm rate.

Although deep learning exhibits strong performance on a wide range of tasks, its intricate models, characterized by a large number of parameters, pose challenges in elucidating their decision-making processes. To address this, the field of explainable artificial intelligence~\cite{ref:Arrieta2019} has surged, aiming to make model decisions transparent and interpretable. Within the context of CNNs, efforts have been made to develop techniques that attempt to understand the decision-making process by reverse mapping the output of the network into the input space to identify the specific input components that were discriminative in producing the output. Class activation mapping (CAM)~\cite{ref:Zhou2015} is one such method, which computes a weighted sum of the outputs of the last convolutional layer using the outputs of the global average pooling layer after the last convolutional layer as weights. It helps identify the regions in the input image that were important for a prediction, but the model needs to be modified to include a global average pooling layer, which may result in lower accuracy. Gradient-weighted class activation mapping (Grad-CAM)~\cite{ref:Selvaraju2017} was introduced as a solution to this limitation of CAM, offering the advantage of not requiring any modifications to the network architecture by using gradient information from the prediction for weighted parameters. Subsequently, Grad-CAM++ ~\cite{ref:Chattopadhay2018}, a generalization of Grad-CAM, and Score-CAM~\cite{ref:Wang2020}, a gradient-free CAM method, were developed to generate more accurate saliency maps than Grad-CAM. These techniques to analyze deep learning models are commonly used in fields such as electrocardiogram signal analysis~\cite{ref:Jahmunah2022} and x-ray diagnosis~\cite{ref:Panwar2020}; however, for GW analysis, they have only been used in Ref.~\cite{ref:Abbott2022} to the best of our knowledge.

In this study, we first take an approach similar to Ref.~\cite{ref:Iess2023} and train a two-dimensional CNN model to classify CCSNe signals using short-time Fourier-transformed spectrograms as input for simplicity. We use nine types of waveforms from recent three-dimensional numerical simulations and O3 real noise to train and validate our model. In the test, signals from sources between 1 and 10 kpc are considered, and the performance of the model for sources at each distance is discussed. To interpret the model, we use three CAM methods to generate saliency maps and evaluate them using two metrics: average drop and average increase. The best CAM method is then applied to correctly classified and also misclassified samples to visualize the regions in the input spectrogram that influence the predictions of our model. 

The remainder of this chapter is organized as follows. Section~\ref{sec:method} describes our data sets, the CNN model, and the CAM techniques. In Sec.~\ref{sec:result}, we discuss the classification performance of our model, and apply multiple visualization techniques to interpret the model. We summarize and conclude the paper in Sec.~\ref{sec:concl}.

\section{Method\label{sec:method}}
Our CNN model is trained to classify strains at the three detectors LIGO Hanford (H1), LIGO Livingston (L1), and Virgo (V1) into ten classes: noise and nine different CCSN waveforms. In this section, we first provide an overview of the data used in this study, including a brief summary of the CCSN simulation data and the preprocessing strategy to generate our training, validation, and test sets. Subsequently, our CNN architecture and the theory of the visualization technique of the model are explained.
\subsection{Data set}
\subsubsection{CCSN waveforms}
Modeling the stellar core collapse, bounce, and subsequent post-bounce evolution is very complicated and computationally expensive. However, remarkable advancements in three-dimensional numerical simulations of neutrino-driven explosions have been achieved by several groups in recent years. Specific details of the waveform depend on various properties of the progenitor, such as the mass, angular velocity, and EOS of the dense matter. Both the general relativity approximation and the handling of neutrino transport critically influence simulations. From the available simulation data under a variety of conditions, we select nine types of waveforms from four recent three-dimensional numerical simulations~\cite{ref:Powell2019, ref:Radice2019, ref:Powell2020, ref:Powell2021}. All of them allow us to compute the GW amplitude in any observer direction from the quadrupole moment.

\citeauthor{ref:Powell2019}~\cite{ref:Powell2019} performed simulations using the general-relativistic neutrino hydrodynamics code CoCoNuT-FMT~\cite{ref:Muller2010}. We use two waveforms from the models \texttt{he3.5} and \texttt{s18}. The progenitor of \texttt{he3.5} is an ultra-stripped star evolved from a helium star with an initial mass of 3.5 $M_\odot$.  The simulation is stopped at 0.7 s after core bounce. The GW is dominated by excitation of $g$-modes in the PNS with a peak frequency around 900 Hz. Model \texttt{s18} is a single star with a zero-age main-sequence (ZAMS) mass of 18 $M_\odot$. The simulation was stopped 0.89 s after core bounce. The GW emission is similar to model \texttt{he3.5}, with $g$-modes oscillations of the PNS with a peak frequency around 900 Hz.

\citeauthor{ref:Radice2019}~\cite{ref:Radice2019} studied eight models using the Eulerian radiation-hydrodynamics code FORNAX~\cite{ref:Skinner2019}. We use waveforms from the models \texttt{s13} and \texttt{s25} corresponding to progenitors of 13 and 25 $M_\odot$ ZAMS, respectively. The simulation is ended at 0.77 s in \texttt{s13} and 0.62 s in \texttt{s25} after bounce. Both waveforms are characterized by $f$- and $g$- modes with a peak frequency around 1400 Hz in \texttt{s13} and 1100 Hz in \texttt{s25}. In addition, the \texttt{s25} waveform has a clear SASI mode around 100 Hz. 

From the simulation by \citeauthor{ref:Powell2020}~\cite{ref:Powell2020}, we use the three models \texttt{m39}, \texttt{y20}, and \texttt{s18np}. Model \texttt{m39} is a rapidly rotating 39 $M_\odot$ Wolf-Rayet star with an initial surface rotation velocity of 600 $\mathrm{km}\ \mathrm{s}^{-1}$. It produces a neutrino-driven explosion without magnetic fields. The other two are nonrotating models of 20 $M_\odot$ Wolf-Rayet star and an 18 $M_\odot$ ZAMS star. The simulation is ended at 0.98, 1.2, 0.56 s after core bounce in models \texttt{m39}, \texttt{y20}, and \texttt{s18np}, respectively. All three models show GW emission associated with prompt convection shortly after bounce as well as $f$-mode oscillations of the PNS. In model \texttt{s18np}, the absence of strong perturbations from convective oxygen burning, in contrast to \texttt{s18}, prevents the shock from being revived and leads to the development of strong SASI activity, with a frequency reaching $\sim$ 400 Hz by the end of the simulation.

\citeauthor{ref:Powell2021}~\cite{ref:Powell2021} performed simulations using three different EOSs: LS220~\cite{ref:Lattimer1991}, SFHx, and SFHo~\cite{ref:Steiner2013}.  The progenitor models are 85 and 100 $M_\odot$ Population III ZAMS stars. We use the \texttt{z85\_sfhx} and \texttt{z100\_sfho} models. The simulation is ended at 0.59 s in \texttt{z85\_sfhx} and 0.62 s in \texttt{z100\_sfho}. Both waveforms show typical $g$-mode emission with a peak frequency of $\sim$ 700 Hz. In the \texttt{z85\_sfhx} model, the frequency of the SASI emission increases up to the point of shock revival, reaching $\sim$ 200 Hz, and decreases afterwards. In the \texttt{z100\_sfho} model, which does not explode, the frequency of the SASI emission continues to increase by the end of the simulation, reaching $\sim$ 400 Hz.

\begin{figure}[tb]
    \centering
    \includegraphics[width=0.98\columnwidth]{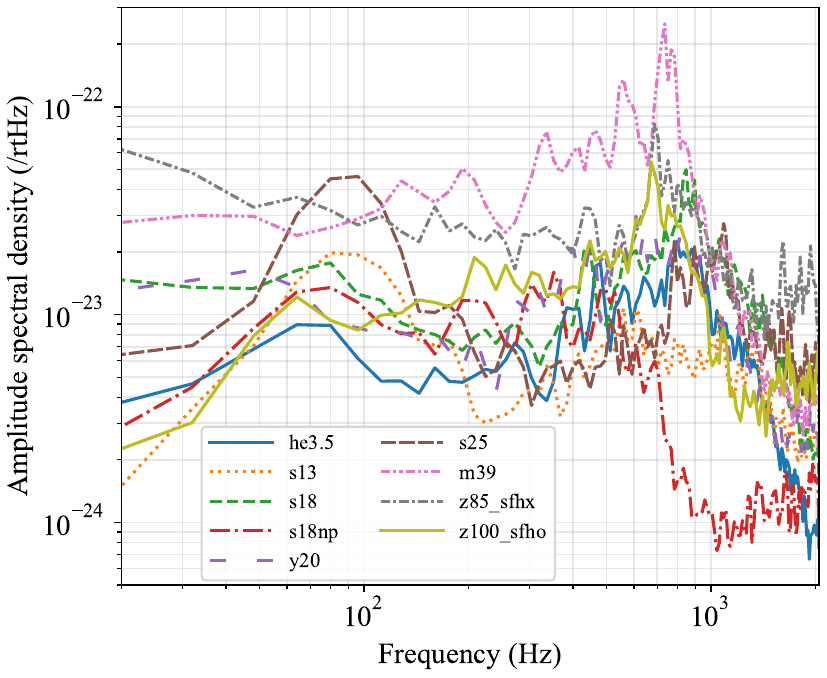}
    \caption{Amplitude spectral density of the plus mode of each waveform at 1 kpc. The observer is in the polar direction.}
    \label{fig:asd}
\end{figure}

Figure~\ref{fig:asd} shows the amplitude spectral density of the plus mode of each waveform at 1 kpc from the polar direction. The amplitude of the \texttt{m39} waveform is the largest of these waveforms. Waveforms that have peaks around 100~Hz such as \texttt{s25} and \texttt{s13} indicate that they have SASI-induced GW modes.

\subsubsection{Data processing}
From the simulation data presented in the previous section, we calculate the amplitude of the GW for generating the data sets. The directions of radiation $(\theta, \phi)$ are uniformly sampled and the plus and cross polarizations of each GW are calculated using the formulas
\begin{align}
    h_+ &= \frac{1}{D}\frac{2G}{c^4}(\ddot{Q}_{\theta\theta}-\ddot{Q}_{\phi\phi}),\\
    h_\times &= \frac{1}{D}\frac{G}{c^4}\ddot{Q}_{\theta\phi},
\end{align}
where $Q$ is the traceless quadrupole moment and $D$ is the distance between a source and Earth. As the sampling of the simulation is usually not uniform in time, we resample data uniformly with a sampling rate of 4096~Hz. A high-pass filter with a cutoff frequency of 11 Hz and a Tukey window with $\alpha = 0.1$ are applied to the resampled signals. Each signal is then truncated or padded with zeros to make the length 1 second. In order to make the model robust, we randomly time shift the signals so that the time of the core bounce is between 0 and 0.15 s. For the training and validation sets, the signals are scaled using an optimal matched filter signal-to-noise  ratio (SNR), as the amplitudes of the simulated signals are quite different. The SNR is defined as
\begin{equation}
    \rho = \sqrt{4\int_{f_{\min}}^{f_{\max}}\frac{|\tilde{h}(f)|^2}{S_n(f)}\mathrm{d}f},
\end{equation}
where $\tilde{h}(f)$ is the Fourier transform of the signal and $S_n(f)$ is the one-side power spectral density of the noise. The network SNR of the detectors H1, L1, and V1, given by
\begin{equation}
    \rho_{\mathrm{net}} = \sqrt{\rho_{\mathrm{H1}}^2+\rho_{\mathrm{L1}}^2+\rho_{\mathrm{V1}}^2},
\end{equation}
is used to scale the signals. We generate samples with network SNRs from 20 to 50 for training and validation sets. For the test set, the signals are scaled to have distances between 1 and 10 kpc. Sky location is also randomly selected and the GW amplitude $h(t)$ is computed, taking into account the antenna pattern functions $F_+$ and $F_\times$ and the delay in arrival time of each detector with the following equation:
\begin{align}
    h(t) &= F_{+}(\alpha, \delta, \psi, t)h_{+}(t+\Delta t)\notag\\
    &\qquad + F_{\times}(\alpha, \delta, \psi, t)h_{\times}(t+\Delta t),
\end{align}
where $\alpha$ is the right ascension, $\delta$ is the declination, and $\psi$ is the polarization angle. $\Delta t$ is the delay in arrival time between the detector and the center of the Earth. We use the PyCBC software library~\cite{ref:Pycbc} to carry out these computations.

Noise used in this study is O3 real data of Advanced LIGO and Advanced Virgo, obtained from the Gravitational Wave Open Science Center~\cite{ref:Abbott2023_a}. Data from GPS time 1238236470 to 1238252308 is used for the training set, 1238265720 to 1238354855 is used for the validation set, and 1238404064 to 1238457121 is used for the test set. Data around the event time reported in the second Gravitational Wave Transient Catalog~\cite{ref:Abbott2021_a} are excluded. After a signal is injected into the noise, each sample is whitened with the power spectral density computed using Welch's method~\cite{ref:Welch1967} and then short-time Fourier transformed with a window size of 0.0625 seconds to produce a spectrogram. The spectrogram is normalized to [0,\ 1] before input to the network.

\begin{figure}[tb]
    \centering
    \includegraphics[width=0.98\columnwidth]{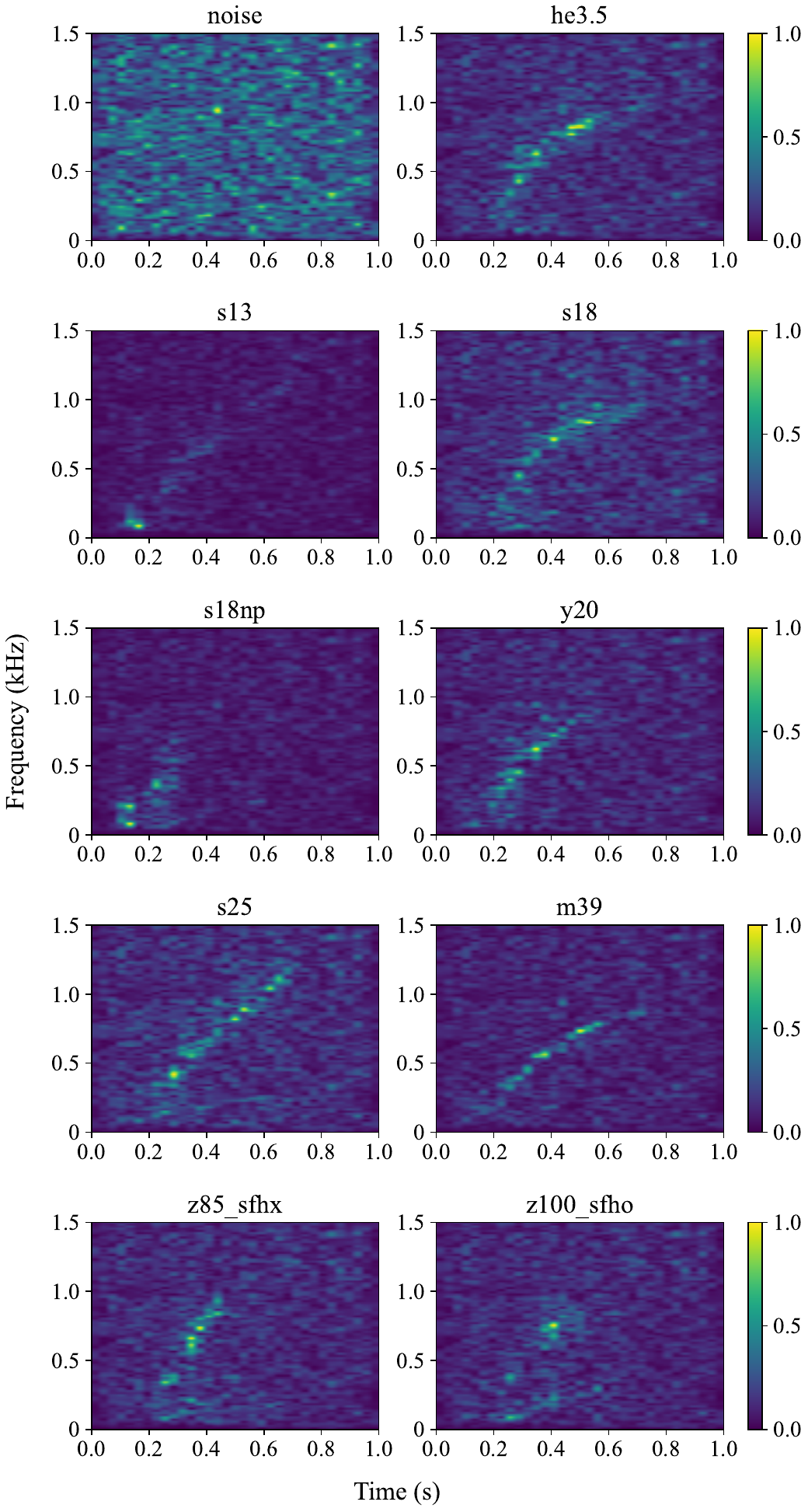}
    \caption{Sample whitened spectrograms of each class at the H1 detector in the training set. Each signal sample is observed in the polar direction and scaled to have an SNR of 40. The bounce time is fixed at 0.1 s.}
    \label{fig:sample}
\end{figure}

We generated 60\ 000 samples for the training and validation sets, and 100\ 000 samples for the test set. The test set has 1000 samples for each class and each distance. Sample spectrograms in the training set are shown in Fig.~\ref{fig:sample}.
\subsection{CNN model}
Our CNN model consists of two convolutional layers of kernel size 3, each followed by a max-pooling layer of size 2 and a rectified linear unit (ReLU) layer. The outputs of these layers are fed into two fully connected layers, and finally the softmax layer outputs a size-10 vector whose elements represent a probability of each class. The model has 427\ 378 trainable parameters in total. This model is shallower than the one used in Ref.~\cite{ref:Iess2023}. However, its classification performance is comparable to the previous study, prompting us to adopt this model. Reducing the number of layers also helps us generate higher-resolution CAM maps.

The model is trained using categorical cross entropy as the loss function and Adam optimizer~\cite{ref:Kingma2014} with a learning rate of $5\times10^{-4}$ to update the weights.  In the training, we adopt curriculum learning~\cite{ref:Bengio2009} as a strategy to enhance the model and accelerate the training by starting from inputting high-SNR samples and gradually adding lower-SNR samples. We train the model on a single GPU (NVIDIA GeForce RTX3090) for 120 epochs with a mini-batch size of 128. 

\subsection{Visualization}
After training the model, we use CAM techniques to generate saliency maps. These maps show the regions in the input that influenced the model's prediction. In this study, we select three CAM methods---Grad-CAM, Grad-CAM++, and Score-CAM---which are widely used today to interpret CNN models. All of these CAM techniques are applied to the convolutional layer prior to the final max-pooling layer in our model.

\subsubsection{Grad-CAM}
Grad-CAM is a gradient-based visualization technique that highlights the important regions of an input image that the model is looking at while making a prediction. Suppose that for a given input, the prediction score for class $c$ before the softmax layer of the trained model is $y^c$, and the $k$th output matrix of the last convolutional layer is $A^k$. To obtain the Grad-CAM map of class $c$, we first compute the gradients of the score $y^c$ with respect to the $(i, j)$ component of the $k$th feature map $A^k$. We then take the global average of these gradients:
\begin{equation}\label{eq:gradcam_weight}
    \alpha_k^c = \frac{1}{Z}\sum_{i, j}\pdv{y^c}{A_{ij}^k},
\end{equation}
where $Z$ is the number of pixels in $A^k$. This weight $\alpha_k^c$ represents the importance of the feature map $k$ for the class $c$.

The Grad-CAM map of the class $c$ is computed as a linear sum of $A^k$ with $\alpha_k^c$ as weights. The ReLU function is applied to extract only features that have a positive contribution to the prediction score. The resulting map of class $c$ is expressed as
\begin{equation}
    L_{\mathrm{Grad-CAM}}^c = \mathrm{ReLU}\pqty{\sum_k \alpha_k^c A^k}.
\end{equation}
Since convolutional layers and pooling layers make the size of the feature map smaller than the input, the Grad-CAM map is finally interpolated to make it the same size as the input.

\subsubsection{Grad-CAM++}
While Grad-CAM takes a global average of the gradient matrix when calculating the weight $\alpha_k^c$ in Eq.~(\ref{eq:gradcam_weight}), \citeauthor{ref:Chattopadhay2018}~\cite{ref:Chattopadhay2018} proposed a method to fully include the importance of each pixel in the gradient matrix by taking its weighted average for the weight:
\begin{equation}\label{eq:gradcampp_weight}
    \alpha_{k}^c=\sum_{i,j}\alpha_{ij}^{kc}\ \mathrm{ReLU}\pqty{\pdv{y^c}{A_{ij}^k}}.
\end{equation}
The ReLU function is used to account for features that increase the activation of the output neuron rather than suppress the activation of the output neuron. The weights $\alpha_{ij}^{kc}$ can be theoretically derived using higher-order derivatives:
\begin{equation}
    \alpha_{ij}^{kc}=\frac{\frac{\partial^2 y^c}{(\partial A_{ij}^k)^2}}{2\frac{\partial^2 y^c}{(\partial A_{ij}^k)^2}+\sum_{a,b}A_{ab}^k\frac{\partial^3 y^c}{(\partial A_{ij}^k)^3}}.
\end{equation}
This method is known as Grad-CAM++, since it can be considered as a generalization of Grad-CAM. The saliency map for Grad-CAM++ is expressed in the same way as for Grad-CAM, using weights in Eq.~(\ref{eq:gradcampp_weight}) and feature maps, as 
\begin{equation}
    L_{\mathrm{Grad-CAM++}}^c = \mathrm{ReLU}\pqty{\sum_k \alpha_k^c A^k}.
\end{equation}
\subsubsection{Score-CAM}
\citeauthor{ref:Wang2020}~\cite{ref:Wang2020} proposed a gradient-free CAM method called Score-CAM. It solves the problem of gradient-based CAM methods, namely that the gradient is unstable, easily disturbed by noise, and can vanish or explode in deep networks. To generate a Score-CAM map, feature maps are used to mask an input image. Let $H^k$ be the $k$th feature map, up-sampled to the same size as the input and normalized to [0, 1]. Given an input image $X$, the weight for the $k$th feature map is computed as the difference between the score of the masked image $X\circ H^k$ and the score of the baseline image $X_\mathrm{b}$:
\begin{equation}
    \alpha_{k}=f(X\circ H^k)-f(X_\mathrm{b}),
\end{equation}
where $f(\cdot)$ denotes the output of the CNN and $\circ$ denotes the Hadamard product. A black image is used as a baseline image. The Score-CAM map of class $c$ is then computed as a linear sum of the $c$th value of $\alpha_k$ and the feature map $A^k$ as 
\begin{equation}
    L_{\mathrm{Score-CAM}}^c = \mathrm{ReLU}\pqty{\sum_k \alpha_k^c A^k}.
\end{equation}
\section{Results and Discussion\label{sec:result}}
\subsection{Classification performance}
\begin{figure}[tb]
    \centering
    \includegraphics[width=0.98\columnwidth]{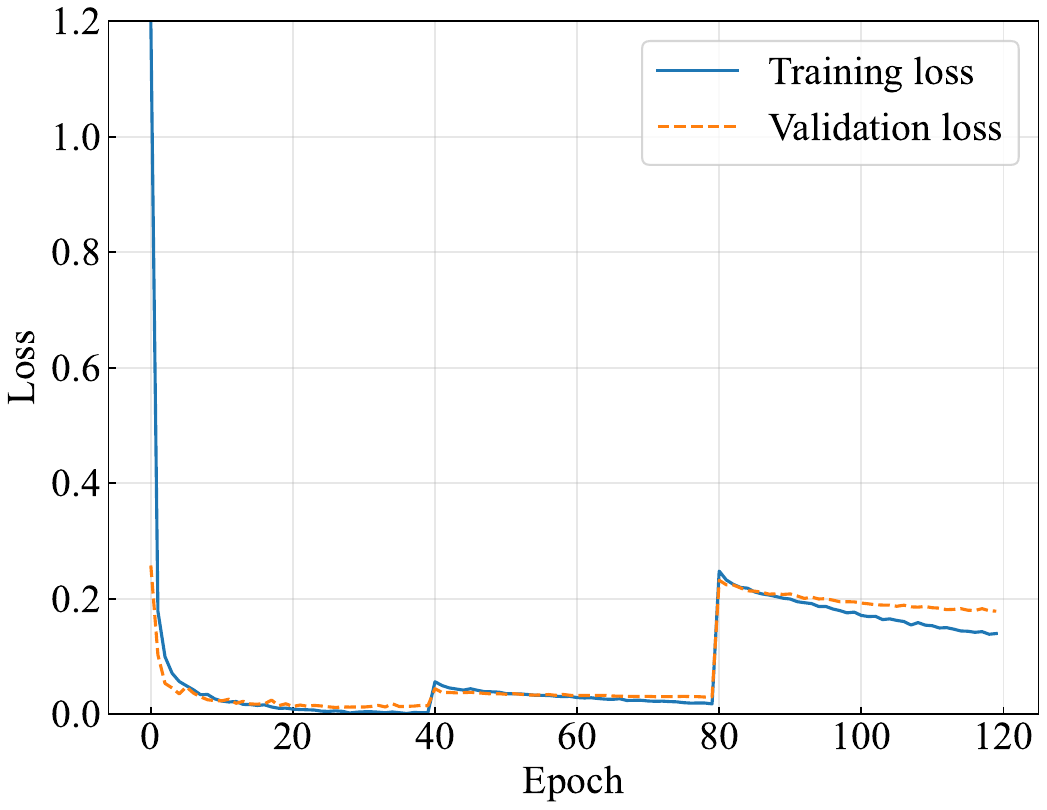}
    \caption{Loss curves for the training and validation sets. During the first 40 epochs, we input samples with SNRs between [40, 50], followed by [30, 40] SNR samples in the subsequent 40 epochs, and [20, 30] SNR samples in the last 40 epochs.}
    \label{fig:loss}
\end{figure}

Figure~\ref{fig:loss} shows the evolution of the categorical cross entropy loss function during the training. Initially, we input samples with SNRs between 40 and 50. In the subsequent 40 epochs, samples with SNRs between 30 and 40 are input, leading to the temporary increase in the loss at epoch 40. In the last 40 epochs, as we input samples with SNRs between 20 and 30, a similar temporary increase in loss is observed at epoch 80. From the loss curves, we confirm that there is no significant overfitting to the training data.
\begin{figure}[tb]
    \centering
    \includegraphics[width=0.98\columnwidth]{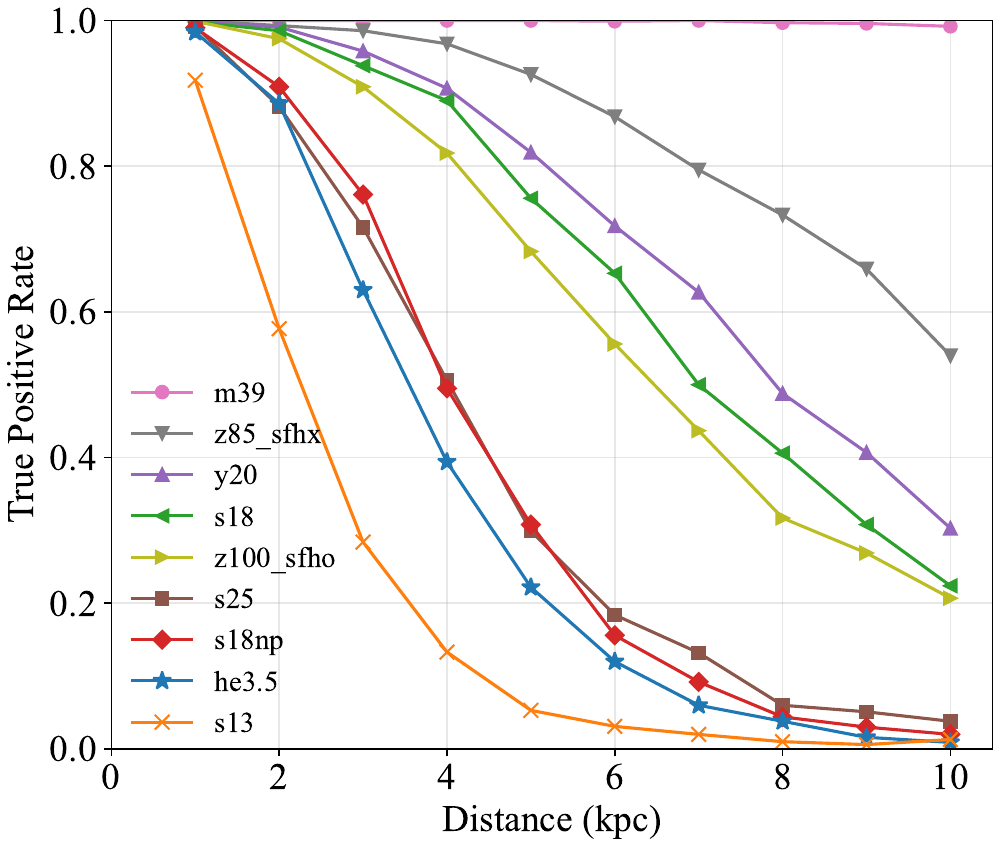}
    \caption{True positive rate of each waveform in the test set against source distance.}
    \label{fig:tpr}
\end{figure}
\begin{figure*}
    \centering
    \begin{tabular}{c}
      \begin{minipage}[t]{0.95\textwidth}
        \centering
        \includegraphics[width=0.95\textwidth]{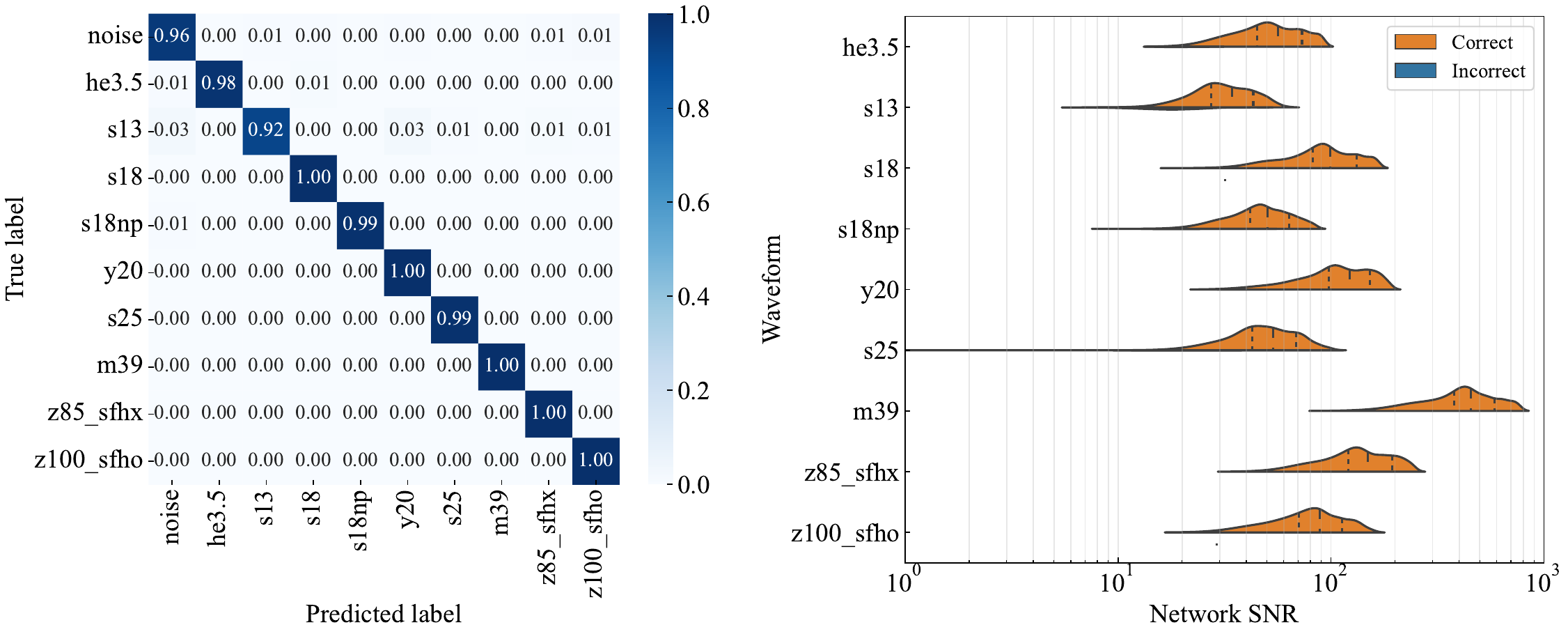}
        \subcaption{1 kpc}
        \vspace{8px}
      \end{minipage} \\
      \begin{minipage}[t]{0.95\textwidth}
        \centering
        \includegraphics[width=0.95\textwidth]{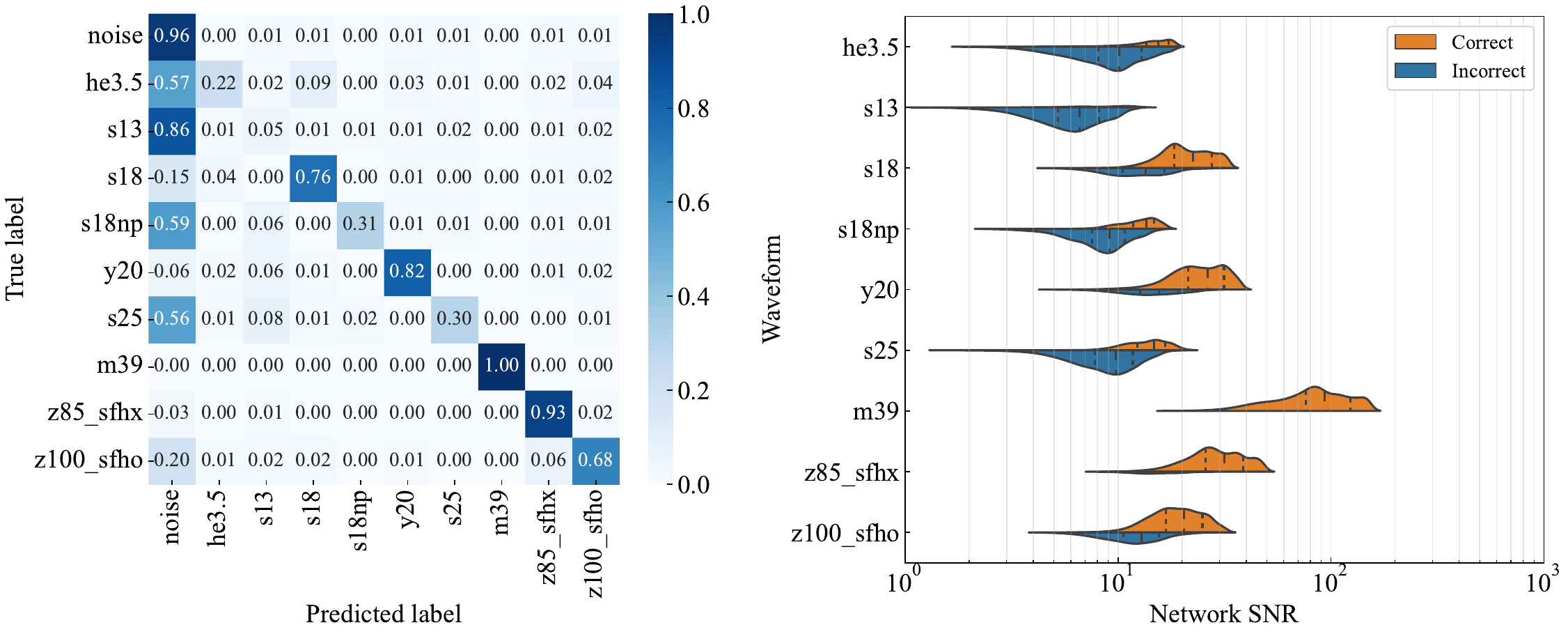}
        \subcaption{5 kpc}
        \vspace{8px}
      \end{minipage} \\
      \begin{minipage}[t]{0.95\textwidth}
        \centering
        \includegraphics[width=0.95\textwidth]{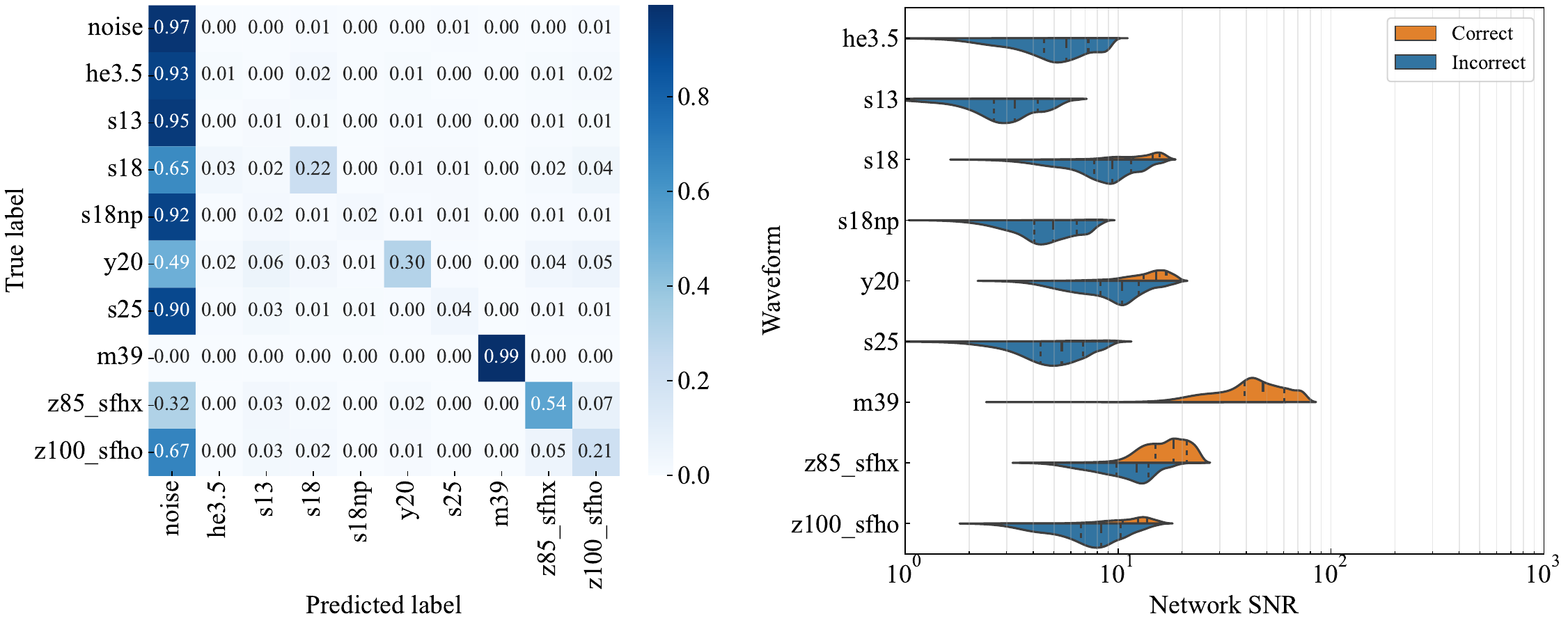}
        \subcaption{10 kpc}
      \end{minipage}
    \end{tabular}
\caption{Confusion matrices of the test set (left) and violin plots of network SNR of each waveform (right) from sources with distances of 1, 5, and 10 kpc.}
\label{fig:cm_violin}
\end{figure*}

Classification accuracy is defined as the proportion of correctly classified samples out of the total number of samples. After training, the model achieves a classification accuracy of 97.8\% on a validation set consisting of uniformly sampled signals with SNRs between 20 and 50. On the test set, our model shows an accuracy of 98.4\% for signals with sources from 1 kpc, which is comparable to the results of the previous study~\cite{ref:Iess2023}, despite some differences in the condition that we used O3 noise instead of O2 noise and performed ten-class classification instead of eight-class classification. In Fig.~\ref{fig:tpr}, we plot a true positive rate (TPR) for each waveform in the test set against distance. A TPR, also known as the sensitivity of a class $c$, is defined as the ratio of the number of samples correctly classified into class $c$ to the number of samples of class $c$ in the test set. For signals from sources at 1 kpc, each waveform has a TPR greater than 90\%, and this decreases monotonically with the distance of the source, having an average TPR of 26.1\% at 10 kpc. For the \texttt{m39} waveform, because the amplitude of the strain is much larger than the others due to its rapid rotation and high explosion energy, the TPR for sources at 10 kpc is 99.2\%.

The performance of a multiclass classifier is also 
expressed by a confusion matrix, which shows the number of samples classified into each class. Figure~\ref{fig:cm_violin} plots the confusion matrices normalized for each class and the distribution of the network SNR for signals from sources at 1, 5, and 10 kpc. We can see from the confusion matrices that as the distance increases, the amplitude of the signal becomes smaller and the number of samples misclassified as noise increases. The accuracy for signals at 10 kpc is 33.2\%, and our model cannot identify most of these signals, except for the \texttt{m39} waveforms, whose SNR is much higher than others with a median value of 47.9. 

\subsection{Dimensionality reduction}
\begin{figure}[tb]
    \centering
    \includegraphics[width=0.98\columnwidth]{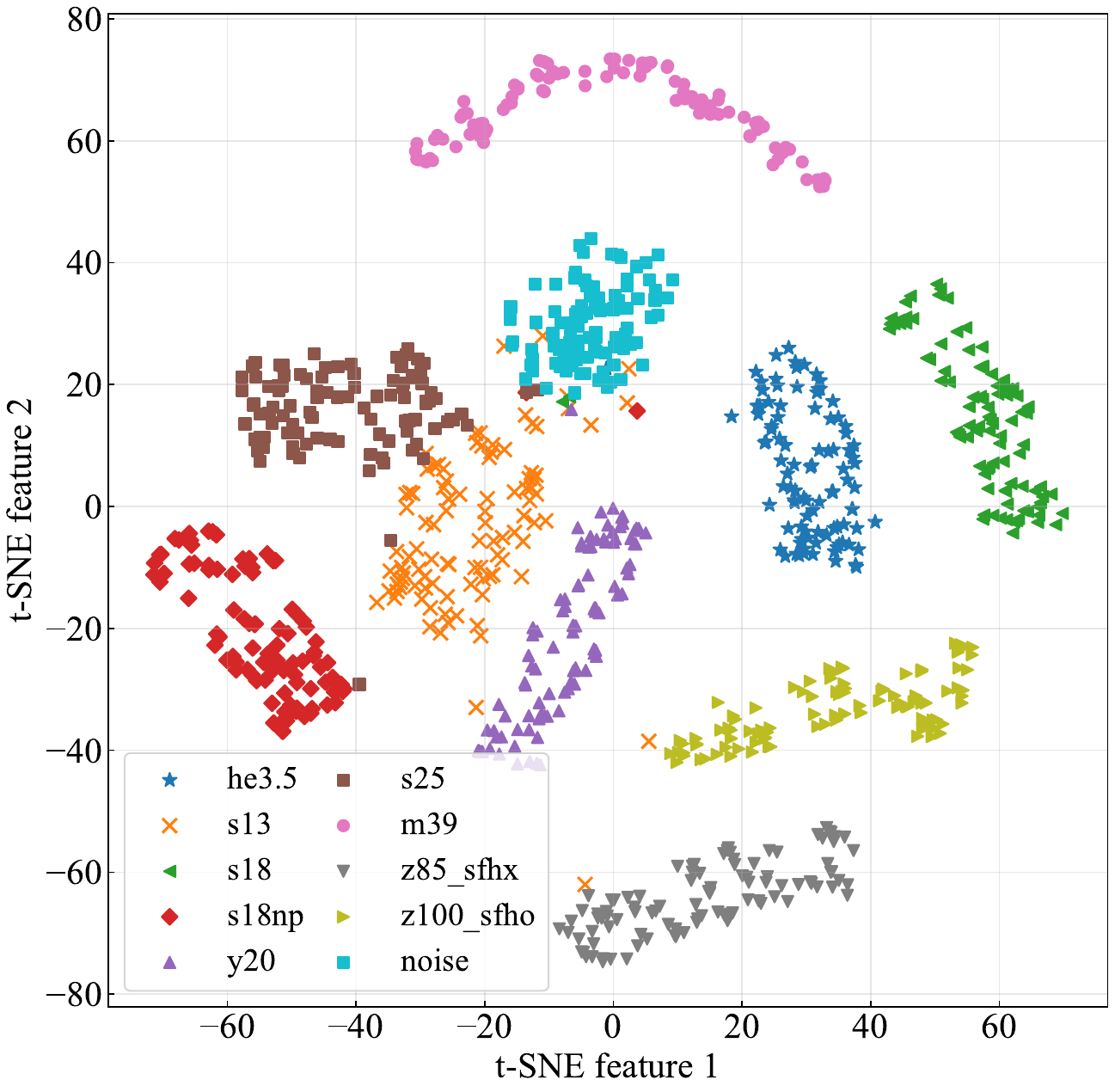}
    \caption{Features of the test samples at 1 kpc extracted by CNN and mapped into two-dimensional space by the t-SNE algorithm.}
    \label{fig:t_sne}
\end{figure}

Before implementing CAM techniques, we used the t-distributed stochastic neighbor embedding (t-SNE)~\cite{ref:van2008} algorithm to see if the convolutional layers in the model can extract the features in the input to classify samples. The t-SNE algorithm is a dimensionality-reduction technique that minimizes the Kullback-Leibler divergence between two probability distributions: one representing pairwise similarities between data points in the original high-dimensional space and another representing pairwise similarities in a lower-dimensional space. In our CNN model, each sample is compressed into a vector with a length of 2112 before the dense layers. The t-SNE algorithm is used to map this vector into two-dimensional space to make it interpretable for humans. We visualize the dimensionally reduced feature maps of the test set, whose signals are coming from sources at 1 kpc, for which our model shows a good classification accuracy. The visualized data are shown in Fig.~\ref{fig:t_sne}. We can clearly see that there are ten clusters in the data set and our model could extract meaningful features to classify these samples into ten classes. The fact that some signal samples are also found in the noise cluster and that \texttt{s13} samples are found in other clusters, especially in the noise cluster, is consistent with the results of the confusion matrix in Fig.~\ref{fig:cm_violin}(a).
\subsection{Saliency maps}
To quantitatively evaluate different CAM methods, we use two metrics---average drop and average increase~\cite{ref:Chattopadhay2018}---which focus on the change in a model's score caused by the explanation map. An explanation map for a target class $c$ is generated as element-wise multiplication of a saliency map $L^c$ with an original image $X$:
\begin{equation}
    E^c = L^c \circ X.
\end{equation}

Average drop measures the percentage decrease in a model's score for a target class $c$ when inputting only the explanation map, instead of the original image. It is expressed as
\begin{equation}
    \mathrm{Average\ drop} = 100\cdot\frac{1}{N}\sum_i \frac{\max(0,y_i^c-o_i^c)}{y_i^c},
\end{equation}
where $y_i^c$ is the score for class $c$ on the $i$th original image and $o_i^c$ is the score on the explanation map. The lower this value, the more effective the visualization method, since the explanation map includes more of  the relevant information for making a correct prediction.

Average increase measures the number of samples in the data set, and the model’s confidence increases when providing only the explanation map as input. It is expressed as
\begin{equation}
    \mathrm{Average\ increase} = 100\cdot\frac{1}{N}\sum_i \Theta(o_i^c-y_i^c),
\end{equation}
where $\Theta$ is the Heaviside step function. Unlike the previous metric, the higher this value is, the more effective the visualization method will be because there are more samples that score higher when given the explanation map than when given the original image.

\begin{figure*}
    \centering
    \includegraphics[width=0.98\textwidth]{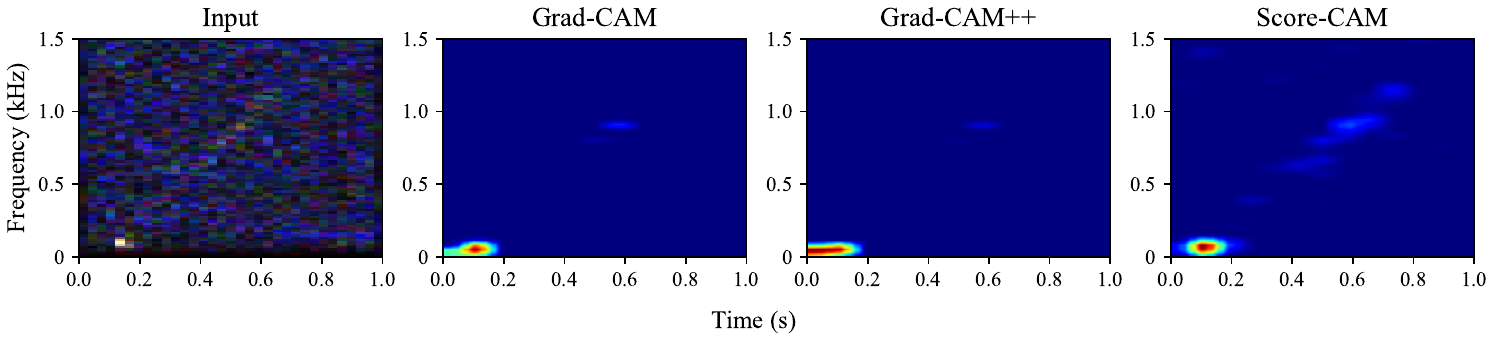}
    \caption{Qualitative comparison of three CAM maps for the \texttt{s25} sample at 1\ kpc.}
    \label{fig:cam_compare}
\end{figure*}

For the three visualization methods Grad-CAM, Grad-CAM++, and Score-CAM, the two metrics described above are computed using signals from sources at 1 kpc in the test set. The results are summarized in Table~\ref{tab:cam_eval}. Score-CAM shows the best results in both metrics, meaning that it is the best visualization technique for our model among the three CAM methods considered in this study. We also qualitatively compare these methods by visualizing some samples. One example is shown in Fig.~\ref{fig:cam_compare}. The input image is represented by a color image, with the red, green, and blue channels corresponding to the H1, L1, and V1 spectrograms, respectively. All three saliency maps take large values around the SASI mode around 100 Hz. At high frequencies, the Grad-CAM and Grad-CAM++ maps only take slightly larger values around 1 kHz, whereas Score-CAM has $g$-mode-like arch shapes around 1 kHz. This suggests that the visualization by Score-CAM captures more of the input features that are discriminative for the prediction.

\begin{table}[tb]
    \caption{Results for evaluation of the explanations generated by Grad-CAM, Grad-CAM++, and Score-CAM on the test set.}
    \label{tab:cam_eval}
    \begin{ruledtabular}
    \begin{tabular}{lccc}
    Method  & Grad-CAM & Grad-CAM++ & Score-CAM \\ \hline
    \begin{tabular}{l}Ave. Drop (\%)\\ (\emph{Lower is better})\end{tabular} & 30.70 & 17.58 & \textbf{9.61} \\ \hline
    \begin{tabular}{l}Ave. Increase (\%)\\ (\emph{Higher is better})\end{tabular} & 1.30 & 1.40 & \textbf{1.96}
    \end{tabular}
    \end{ruledtabular}
\end{table}
\begin{figure*}
    \centering
    \includegraphics[width=.98\textwidth]{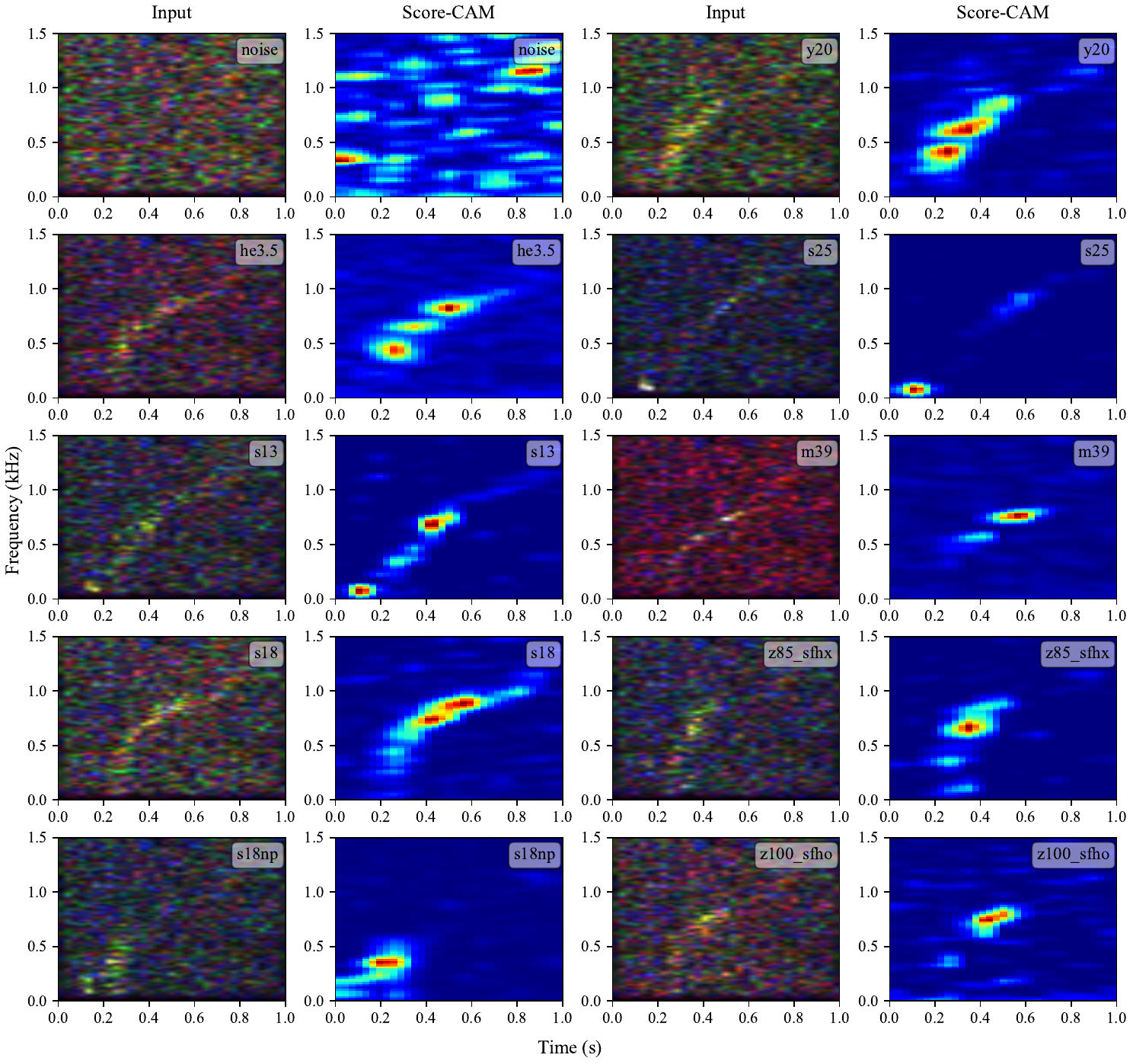}
    \caption{Input spectrograms and Score-CAM maps of correctly classified samples. The SNR of each signal sample is 40.}
    \label{fig:cam}
\end{figure*}

As discussed above, we determine that the Score-CAM is the optimal method for generating saliency maps for our model. We produce saliency maps by Score-CAM for the inputs of each class, which can be seen in Fig.~\ref{fig:cam}. In the input images, as in the previous figure, the red, green, and blue channels corresponding to the H1, L1, and V1 data, respectively. This means, for example, that in the reddish image such as the \texttt{m39} sample in Fig.~\ref{fig:cam}, the SNR at the H1 detector is smaller than that at the L1 and V1 detectors. All of the plotted signal samples are scaled to have an SNR of 40 and are correctly classified by our model. We plotted several CAM maps for noise samples in addition to the one in this figure, but the regions that the model sees to predict them to be noise were random. In the \texttt{he3.5} and \texttt{s13} samples, we can see that the model focuses on the $g$-mode arch shape, especially in their low- and high- frequencies areas. In the \texttt{s18} and \texttt{y20} models, the models see all of the $g$-modes. In the \texttt{s18np} model, the CAM map indicates that the model considers not only the $g$-mode but also prompt convection and SASI. The \texttt{s25} model has SASI activity, but its amplitude is not too large, and the CAM map shows that the model's prediction is based on the prompt convection and the high-frequency $g$-mode. In the \texttt{m39}, \texttt{z85\_sfhx}, and \texttt{z100\_sfho} models, the CAM maps take large values at high frequencies in the $g$-mode. In addition, in the \texttt{z85\_sfhx} and \texttt{z100\_sfho} models, the low-frequency SASI mode, whose frequency increases with time, is also visible in the CAM maps. To summarize these outcomes, we found that the model looks at the $g$-mode in all signal waveforms, and also looks at SASI and prompt convection in some signal waveforms when classified.

\begin{figure*}[tb]
  \begin{minipage}[t]{0.32\textwidth}
    \centering
    \includegraphics[width=0.98\textwidth]{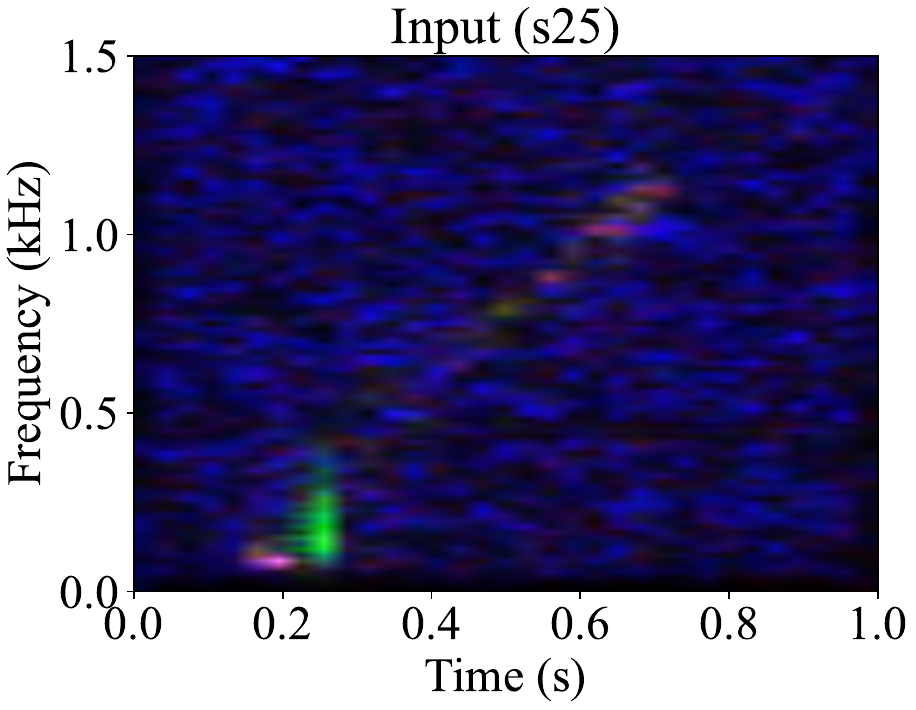}
    \subcaption{}
  \end{minipage}
  \begin{minipage}[t]{0.32\textwidth}
    \centering
    \includegraphics[width=0.98\textwidth]{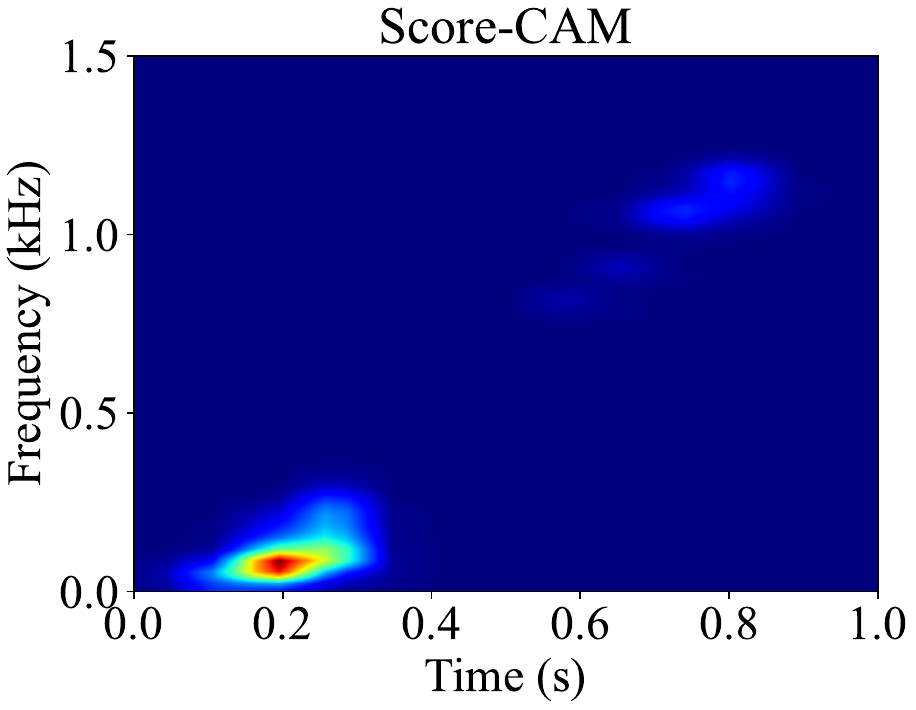}
    \subcaption{}
  \end{minipage}
  \begin{minipage}[t]{0.32\textwidth}
    \centering
    \includegraphics[width=0.98\textwidth]{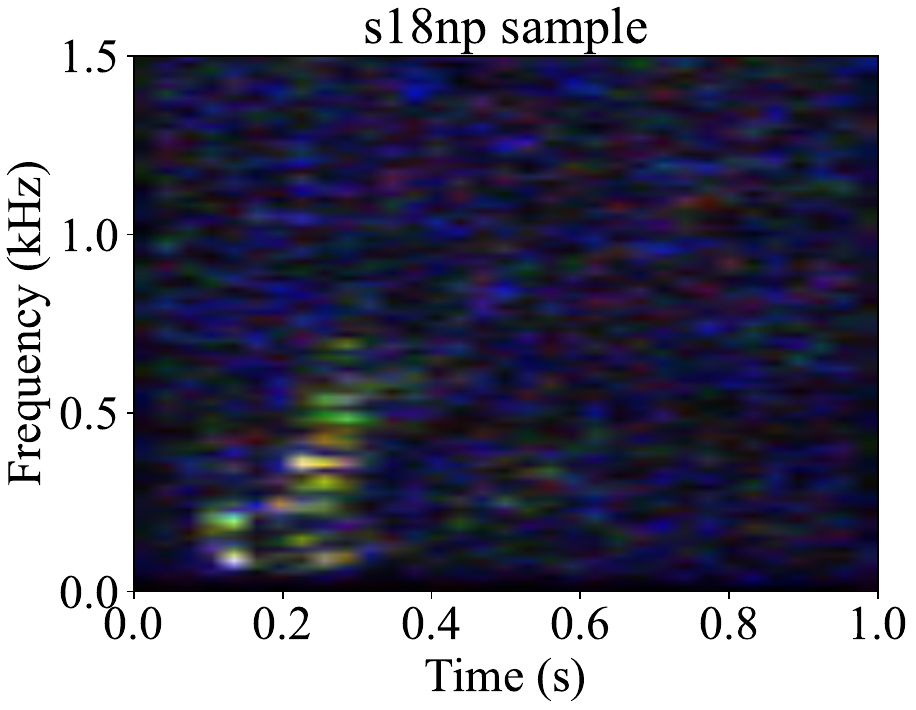}
    \subcaption{}
  \end{minipage}
  \caption{\texttt{s25} sample classified as \texttt{s18np}. (a) Input spectrogram. The red, green, and blue channels correspond to the H1, L1, and V1 data, respectively. There is a glitch in the L1 data. (b) Score-CAM map. (c) Example spectrogram of a \texttt{s18np} sample.}
  \label{fig:mis_sample1}
\end{figure*}

\begin{figure*}[tb]
  \begin{minipage}[t]{0.32\textwidth}
    \centering
    \includegraphics[width=0.98\textwidth]{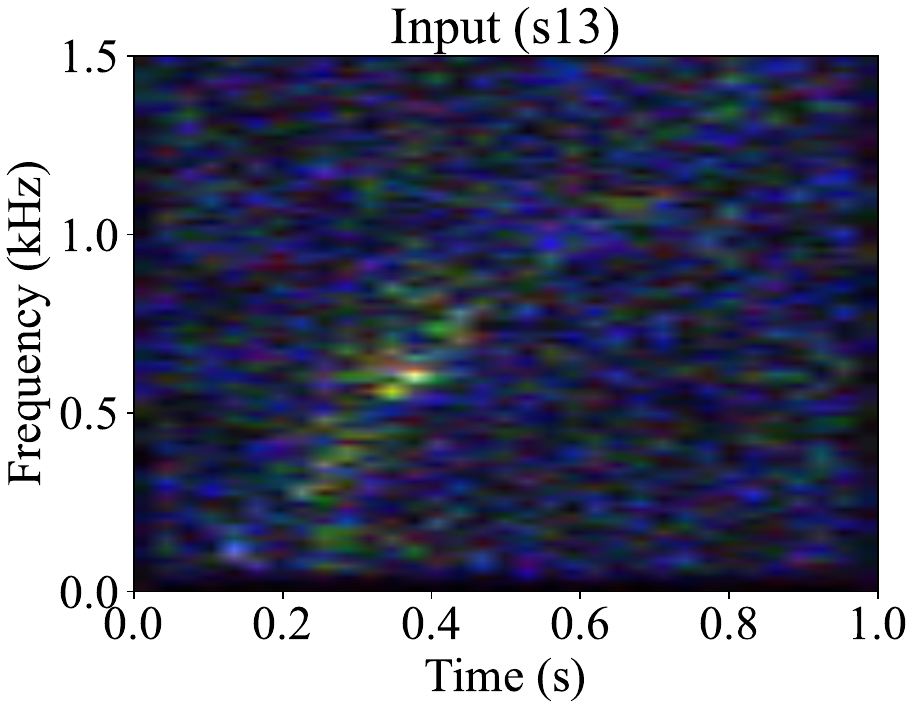}
    \subcaption{}
  \end{minipage}
  \begin{minipage}[t]{0.32\textwidth}
    \centering
    \includegraphics[width=0.98\textwidth]{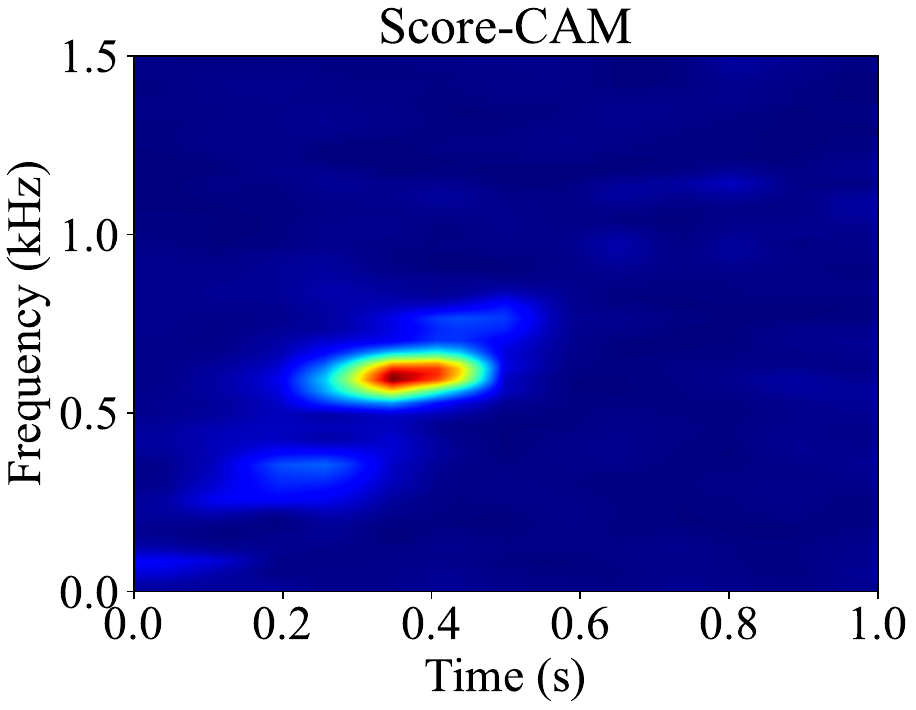}
    \subcaption{}
  \end{minipage}
  \begin{minipage}[t]{0.32\textwidth}
    \centering
    \includegraphics[width=0.98\textwidth]{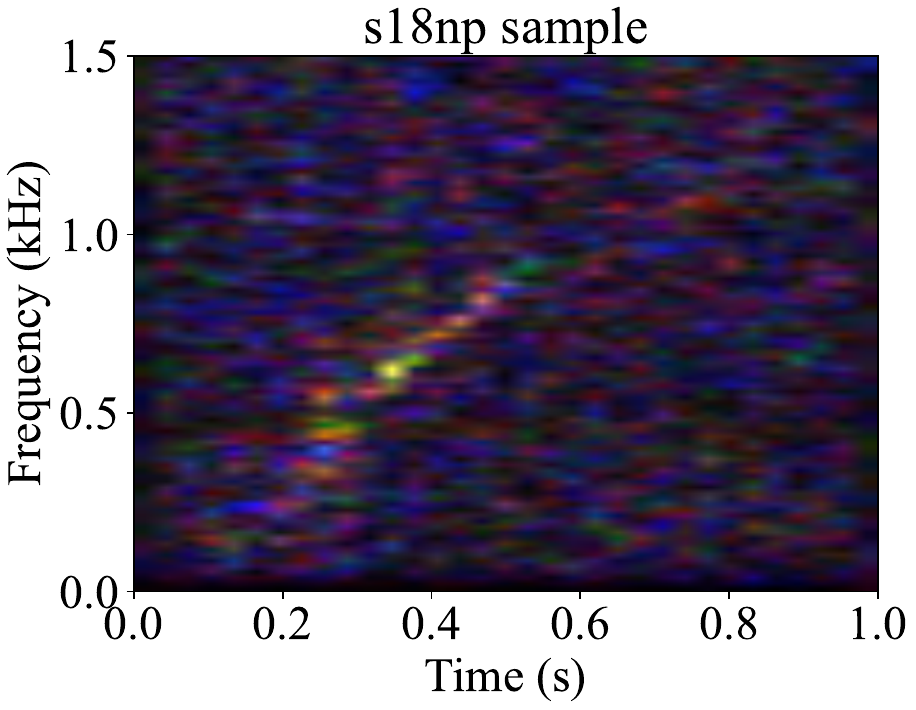}
    \subcaption{}
  \end{minipage}
  \caption{\texttt{s13} sample classified as \texttt{y20}. (a) Input spectrogram. The red, green, and blue channels correspond to the H1, L1, and V1 data, respectively. (b) Score-CAM map. (c) Example spectrogram of a \texttt{y20} sample.}
  \label{fig:mis_sample2}
\end{figure*}

Additionally, we plot saliency maps of the misclassified samples. Figure~\ref{fig:mis_sample1} shows a spectrogram of the \texttt{s25} signal sample and the Score-CAM map, which the model classified as \texttt{s18np}. An example spectrogram of the correct class \texttt{s18np} is also shown. The SNR of this signal is 85, which is quite large, and the $g$-mode and the prompt convection are visible, but there is a glitch in the strain at the L1 detector. The Score-CAM map shows that the model focuses on the prompt convection and the glitch, which are used to determine that the signal is \texttt{s18np}. Because of this glitch, the model predicted the signal as \texttt{s18np}, whose $g$-mode frequency increases in a shorter period of time. Another example is plotted in Fig.~\ref{fig:mis_sample2}. This sample contains a \texttt{s13} signal with an SNR of 48, and there are no glitches, but the model classified it as \texttt{y20}. We can see a SASI-induced GW mode around 100 Hz from 0.2 to 0.4 s, but the Score-CAM map indicates that the model only looks at a portion of the $g$-mode and does not see the low-frequency mode. 

From the misclassified samples and the Score-CAM maps, it is found that the performance of the model is sometimes affected by glitches, and does not fully take advantage of the characteristics of the signals. The former could be resolved by generating training sets that contain more glitches, and the latter could be resolved by using a time-frequency representation that is better able to reflect the various features of the CCSN signals.

\section{Conclusions\label{sec:concl}}
In this study, we trained a two-dimensional CNN model to classify CCSN GW signals immersed in real noise of O3 observation data. Our model showed a comparable result to the previous study~\cite{ref:Iess2023} for signals from sources with distances of 1 kpc. To interpret the trained model, we used the t-SNE algorithm and mapped the extracted features by the convolutional layers into a two-dimensional space. The dimension-reduced features show that the convolutional filters could extract meaningful features that are significant for classifying the signals. To gain insights into the decision-making process of the model, we applied the CAM technique to visualize the regions in the inputs that were influential to the predictions. Three methods---Grad-CAM, Grad-CAM++, and Score-CAM---were considered and we concluded that Score-CAM is the best for our model in terms of the average drop and average increase metrics. The Score-CAM maps of correctly classified signal samples revealed that the model's predictions were heavily affected by a part of the entire $g$-mode in the spectrogram of each signal. In some waveform models such as \texttt{s18np}, \texttt{s25}, \texttt{z85\_sfho}, and \texttt{z100\_sfho}, their CAM maps suggest that the prompt convection or SASI-induced GW mode also affects the model's prediction.

It is important to note that $\sim$ 4\% of the pure noise test samples are identified as signal, which means that our model produces a false alarm every $\sim$ 25 s, making it unsuitable as a detection pipeline. Since this study is the first to focus on the interpretability of CNN models in GW data analysis and serves as a first step in showcasing the effectiveness of the CAM techniques, we did not prioritize its viability as a detection methodology. To utilize machine learning models for a future detection pipeline, it is crucial to lower the false alarm rate. Using the CAM techniques can potentially enhance the efficacy of CNN models for this purpose.

In this analysis, a time-frequency map was created from the short-time Fourier transform, but its resolution is limited by the uncertainty relationship between time and frequency. In future studies, we would like to improve the accuracy of the CNN model by using methods such as the Hilbert-Huang transform~\cite{ref:Takeda2021}, which can generate higher-resolution time-frequency maps, and to confirm that the CNN can also utilize several more GW modes to classify CCSN signals.

\begin{acknowledgments}
The authors would like to thank Jade Powell for providing gravitational-wave simulation data.
This research was supported in part by JSPS Grant-in-Aid for Scientific Research [No. 22H01228 (K.\ Somiya), and Nos. 19H01901, 23H01176 and 23H04520 (H.\ Takahashi)].
This research was also supported by the Joint Research Program of the Institute for Cosmic Ray Research, University of Tokyo and Tokyo City University Prioritized Studies.
This research has made use of data or software obtained from the Gravitational Wave Open Science Center (gwosc.org), a service of the LIGO Scientific Collaboration, the Virgo Collaboration, and KAGRA. This material is based upon work supported by NSF's LIGO Laboratory which is a major facility fully funded by the National Science Foundation, as well as the Science and Technology Facilities Council (STFC) of the United Kingdom, the Max-Planck-Society (MPS), and the State of Niedersachsen/Germany for support of the construction of Advanced LIGO and construction and operation of the GEO600 detector. Additional support for Advanced LIGO was provided by the Australian Research Council. Virgo is funded, through the European Gravitational Observatory (EGO), by the French Centre National de Recherche Scientifique (CNRS), the Italian Istituto Nazionale di Fisica Nucleare (INFN) and the Dutch Nikhef, with contributions by institutions from Belgium, Germany, Greece, Hungary, Ireland, Japan, Monaco, Poland, Portugal, Spain. KAGRA is supported by Ministry of Education, Culture, Sports, Science and Technology (MEXT), Japan Society for the Promotion of Science (JSPS) in Japan; National Research Foundation (NRF) and Ministry of Science and ICT (MSIT) in Korea; Academia Sinica (AS) and National Science and Technology Council (NSTC) in Taiwan. 
\end{acknowledgments}
\bibliography{apssamp}

\end{document}